\documentclass[prd,nofootinbib,twocolumn,superscriptaddress,preprintnumbers,balancelastpage,nofootinbib]{revtex4-1}

\usepackage{graphicx}  
\usepackage{dcolumn}   
\usepackage{bm}        
\usepackage{amssymb, amsmath}
\usepackage{slashed}   
\usepackage{hyperref}

\usepackage[usenames,dvipsnames,svgnames,table]{xcolor}

\newcommand{\cor}[1]{{\color{Red}}}

\def\be{\begin{equation}}
\def\ee{\end{equation}}
\def\bea{\begin{eqnarray}}
\def\eea{\end{eqnarray}}

\newcommand{\apr}{{A^\prime}}
\newcommand{\vmin}{v_{\rm min}}

\newcommand{\vecq}{\mathbf{q}}

\newcommand{\GeV}{{\rm GeV}}
\newcommand{\MeV}{{\rm MeV}}
\newcommand{\keV}{{\rm keV}}
\newcommand{\eV}{{\rm eV}}

\def\beq{\begin{equation}}
\def\eeq{\end{equation}}

\def\be{\begin{eqnarray}}
\def\ee{\end{eqnarray}}

\begin{document}

\newcommand\FNAL{Fermi National Accelerator Laboratory, Batavia, IL USA }
\newcommand\KICP{Kavli Institute for Cosmological Physics, University of Chicago,  Chicago, IL USA}
\newcommand\EFI{Enrico Fermi Institute, University of Chicago, Chicago, IL USA}
\newcommand\UIUC{University of Illinois at Urbana-Champaign, Urbana, IL USA}

\preprint{FERMILAB-PUB-19-257-A}

\title{Electron Ionization via Dark Matter-Electron Scattering and the Migdal Effect}

\author{Daniel Baxter}
\email{dbaxter@kicp.uchicago.edu}
\affiliation{\KICP}
\affiliation{\EFI}
\author{Yonatan Kahn}
\email{yfkahn@illinois.edu}
\affiliation{\KICP}
\affiliation{\UIUC}
\author{Gordan Krnjaic}
\email{krnjaicg@fnal.gov}
\affiliation{\FNAL}

\date{\today}

\begin{abstract}
\noindent
There are currently several existing and proposed experiments designed to probe sub-GeV dark matter (DM) using electron ionization in various materials. The projected signal rates for these experiments assume that this ionization yield arises only from DM scattering directly off electron targets, ignoring secondary ionization contributions from DM scattering off nuclear targets. We investigate the validity of this assumption and show that if sub-GeV DM couples with comparable strength to both protons and electrons, as would be the case for a dark photon mediator, the ionization signal from atomic scattering via the Migdal effect scales with the atomic number $Z$ and 3-momentum transfer $\vecq$ as $Z^2  \mathbf{q}^2$. The result is that the Migdal effect is always subdominant to electron scattering when the mediator is light, but that Migdal-induced ionization can dominate over electron scattering for heavy mediators and DM masses in the hundreds of MeV range. We put these two ionization processes on identical theoretical footing, address some theoretical uncertainties in the choice of atomic wavefunctions used to compute rates, and discuss the implications for DM scenarios where the Migdal process dominates, including for XENON10, XENON100, and the recent XENON1T results on light DM scattering.
\noindent
\end{abstract}

\maketitle




\section{Introduction}
Although the evidence for dark matter (DM) is overwhelming, its microscopic properties remain unknown and motivate
 various experimental techniques to uncover its possible non-gravitational interactions \cite{Bertone:2016nfn}. 
 In recent years, there have been several novel experimental techniques introduced to directly detect 
 hitherto inaccessible DM candidates below the GeV scale~\cite{Battaglieri:2017aum}.  One particularly promising strategy involves 
 detecting single electron ionization from DM-electron scattering \cite{Essig:2011nj,Essig:2012yx,Graham:2012su,An:2014twa,Lee:2015qva,Essig:2015cda,Hochberg:2015pha,Hochberg:2015fth,Cavoto:2016lqo,Derenzo:2016fse,Essig:2016crl,Hochberg:2016ntt,Kouvaris:2016afs,Essig:2017kqs,Cavoto:2017otc,Fichet:2017bng,Tiffenberg:2017aac,Hochberg:2017wce,Romani:2017iwi,Agnes:2018oej,Crisler:2018gci,Agnese:2018col,Essig:2018tss,Abramoff:2019dfb,Emken:2019tni}. 
 
 Since momentum transfer from the DM to a target particle $T$ is most efficient when $m_{\rm DM} > m_T$, a bound atomic electron can capture an order-1 fraction of the DM kinetic energy for $m_{\rm DM} > m_e$ and be ionized. Similar reasoning would suggest that DM lighter than an atomic nucleus cannot efficiently transfer momentum to the nucleus, which is why experiments searching for nuclear recoils are typically insensitive to $m_{\rm DM} < m_p$. However, in a bound atomic system, momentum transfer to the \emph{entire atom} will be redistributed among electrons and the nucleus through the electronic binding energy. This is known as the Migdal effect \cite{Migdal,Vergados:2004bm,Moustakidis:2005gx,Bernabei:2007jz,Ibe:2017yqa,Dolan:2017xbu,Bell:2019egg} and can also result in a final state with an ionized electron and a recoiling atom. Until now, the Migdal effect has solely been used to set limits on nucleon coupling for low mass weakly interacting massive particle models~\cite{Akerib:2018hck,armengaud:2019edel,liu:2019cdex,Aprile:2019mig}.
    
The main result of this paper is the following: in models where sub-GeV DM couples comparably to electrons and protons, the ratio of the differential ionization rate $dR_M/d\vecq$ due to the Migdal effect (which we will refer to as ``Migdal scattering" for brevity) to the corresponding direct electron scattering rate $dR_e/d\vecq$ satisfies
\be
\label{eq:money}
\frac{dR_M/d\vecq}{dR_e/d\vecq} >  Z^2  \left(\frac{m_e}{m_N}\right)^2 (\vecq r_a)^2,
\ee
 where $m_N$ is the mass of the target nucleus, $Z$ its atomic number, $\vecq$ the 3-momentum transferred from the DM to the atom, and $r_a$ is an effective atomic radius which we will define more precisely in Sec.~\ref{sec:Dynamics}.\footnote{If DM couples equally to protons and neutrons, $Z$ should be replaced by the mass number $A$.}
 As we will show, the rate computation for both electron scattering and Migdal scattering involves identical atomic ingredients because the initial hard scatter factorizes from the dynamics of ionization in bound atoms. However, the scattering probability in the latter case is enhanced by $Z^2$ due to coherent scattering off the nucleus, and simultaneously suppressed by the small electron mass compared to the heavy nucleus, though this suppression is mitigated somewhat when the momentum transfer to the atom is large. 
 
 Due to these competing effects, the Migdal scattering rate in heavy atoms such as Xe is generically dominated by the \emph{largest} kinematically-permitted momentum transfers, typically hundreds of keV, which are small on nuclear scales but large on electron scales; by contrast, direct DM-electron scattering is dominated by the {\it smallest} momentum transfers. Thus, the direct DM-electron scattering rate generically dominates over Migdal scattering for light-mediator exchange, which favors small momentum transfers, whereas Migdal scattering can dominate for heavier mediators and heavier DM which imparts larger momentum transfers to the target system. When Eq.~(\ref{eq:money}) is integrated over the momentum transfer $\vecq$, the total rate $R_M$ will then dominate over $R_e$ for sufficiently heavy DM.

This paper is organized as follows. In Sec.~\ref{sec:RefModel} we define our reference model of DM coupling to both electrons and protons. In Sec.~\ref{sec:spectra_derivation} we develop in parallel the formalisms for electron and Migdal scattering, illustrating their similarities and differences. In Sec.~\ref{sec:ion} we discuss the conversion from electron recoil spectra to observed ionization spectra in Xe and highlight the importance of the electron binding energies and wavefunctions in obtaining accurate exclusion curves. We conclude in Sec.~\ref{sec:conclusion} with a comparison of Migdal and electron exclusion curves for XENON10 and XENON100 data \cite{Angle:2011th,Aprile:2013blg}, as well as the recent XENON1T results on light DM scattering \cite{Aprile:2019xxb}. We emphasize throughout that considerable theoretical uncertainty exists as to the correct choice of wavefunctions to use in computing limits on both Migdal and electron scattering. Consequently, our limits presented here should be considered provisional pending a dedicated analysis of relativistic and electron correlation effects in heavy atomic systems.




\section{Reference Model}
\label{sec:RefModel}
Our benchmark model consists of a DM candidate $\chi$, which scatters off both 
electrons and protons through the exchange of a massive dark photon $\apr$ \cite{Holdom:1985ag,Okun:1982xi}. The 
$\apr$ kinetically mixes with the visible photon, and after rotating away the kinetic mixing term $\frac{\epsilon}{2} F^{\mu\nu} F^\prime_{\mu\nu}$, the Lagrangian for this scenario contains 
\be
{\cal L} = - \frac{1}{4} F^\prime_{\mu \nu}F^{\prime \,\mu \nu} + \frac{m^2_{\apr}}{2} A^\prime_\mu A^{\prime \, \mu} + A^\prime_\mu \left(  g_D J_D^
\mu + \epsilon e J_{\rm EM}^\mu \right),~~~~
\ee
where $\epsilon$ is the kinetic mixing parameter, $m_{\apr}$ is the $\apr$ mass, $J_{\rm EM}^\mu = \sum_f Q_f \bar f \gamma^\mu f$ is the electromagnetic current  of all 
Standard Model fermions $f$ with charges $Q_f$, $g_D~\equiv~\sqrt{4\pi \alpha_D}$ is the dark photon gauge coupling,  and 
$J_{ D}^\mu~= \bar \chi \gamma^\mu \chi$ or $i \chi^* \partial_\mu \chi$ is the DM current for a Dirac fermion or
complex scalar DM candidate, respectively.
The fiducial non-relativistic cross section for  $\chi$ scattering off a
free-particle target $T$ is defined at a reference 3-momentum transfer $\mathbf{q}_0$ as
\be
\label{cross-section}
\overline \sigma_T = \frac{16 \pi \epsilon^2 \alpha \alpha_D \mu^2_{\chi T}}{(m_{A'}^2 + |\mathbf{q}_0|^2)^2},
\ee
where $\mu_{\chi T}$ is the $\chi$-$T$ reduced mass; by coincidence
 this same parametric expression hold for both complex scalar and Dirac fermion DM candidates.

This popular scenario features comparable mediator couplings to electrons and protons, so it
serves as a good benchmark for comparing DM-induced 
ionization from direct electron scattering and Migdal scattering. Indeed, both processes will always
be present, so for the remainder of this paper, we will only consider this model. However,
 comparable quark and lepton couplings are by no means unique to dark photons. This feature applies to most anomaly-free  
 $U(1)$ extensions to the SM (e.g. gauged $B-L$)  whose gauge bosons couple to DM \cite{Bauer:2018onh};
  it is also generic for (pseudo)scalar-mediated DM scattering to feature comparable electron and proton
  couplings \cite{Krnjaic:2015mbs, Knapen:2017xzo}.
 
We note for completeness that, in the general case with arbitrary mediator couplings 
to different particle species, the relationship between
Migdal and electron scattering  in Eq.~(\ref{eq:money}) becomes
\be
\label{eq:moneyALT}
\frac{dR_M/d\vecq}{dR_e/d\vecq} \gtrsim \left(\frac{Z g_p + (A-Z)g_n}{g_e}\right)^2 \left(\frac{m_e}{m_N}\right)^2 (\vecq r_a)^2,~~
\ee 
where $g_e$ and $g_{p,n}$ are the mediator's couplings to electrons, protons, and neutrons respectively. Thus, unless $g_{p,n} \ll g_e$,
there is a kinematic regime for which Migdal scattering can predominate over electron scattering.




\section{Comparison of Electron Scattering and Migdal Scattering}
\label{sec:spectra_derivation}

In this section, we carefully define the kinematics and dynamics relevant for sub-GeV DM interacting with atoms through electron and Migdal scattering. We will work in the framework of the dark photon model described above in Sec.~\ref{sec:RefModel}, where the dark photon mediates DM-SM interactions and couples equally to electrons and protons with strength $\epsilon |e|$, but our results are applicable to any model where the momentum dependence of DM-electron and DM-proton scattering (that is, the form factor $F_{\rm DM}(q)$ defined below Eq.~(\ref{eq:Fdef})) is identical.

\subsection{Kinematics}

In both Migdal and electron scattering, the incoming and outgoing states are the same: a DM particle plus a bound atom and a DM particle plus an ionized atom plus an unbound electron, respectively. The incoming DM is assumed to be a plane wave, which is both an energy eigenstate and a momentum eigenstate. The incoming atom (at rest in the lab frame) is an energy eigenstate, and is also a momentum eigenstate for the \emph{total} momentum of the atom $\mathbf{p}_A =  \mathbf{p}_N + \sum_{i = 1}^{Z} \mathbf{p}_i$, where the sum runs over the $Z$ electrons in the electron cloud of the (neutral) atom. The outgoing DM is also a plane wave, but the outgoing atom can either be treated as an ionized atom with a separate ionized electron, or an atom in an excited state where the ionized electron belongs to the continuum spectrum of the atomic Hamiltonian. Following Ref.~\cite{Ibe:2017yqa}, we will take the second perspective where energy-momentum conservation is more transparent, in which case the entire atom recoils with velocity $\mathbf{v}_A$ and has momentum $\mathbf{p}_A = \overline{m}_A \mathbf{v}_A$, where $\overline{m}_A = m_N + Z m_e$ is the nominal mass of the atom neglecting binding energy. In all cases we will consider, it is appropriate to approximate $\overline{m}_A$ by $m_N$ since the nucleus is so much heavier than the electron cloud. The energetics of the ionized electron are accounted for by treating it as an excited state of the electron cloud.

To summarize, when treating the atom as a composite system of electrons and nucleus with a spectrum of internal energy levels, both energy and momentum are conserved in DM-atom interactions. For DM with mass $m_\chi$, incoming velocity $\mathbf{v}$, and outgoing momentum $\mathbf{p}'_\chi$, momentum
conservation requires
\be
\label{eq:qcons}
\mathbf{q} & \equiv m_\chi \mathbf{v} - \mathbf{p}'_\chi = m_N \mathbf{v}_A 
\ee
and energy conservation requires
\be
\Delta E_e  = \frac{1}{2}m_\chi v^2 - \frac{|m_\chi \mathbf{v} - \mathbf{q}|^2}{2m_\chi} - \frac{\mathbf{q}^2}{2m_N} = \mathbf{q} \cdot \mathbf{v} - \frac{\mathbf{q}^2}{2 \mu_{\chi N}} ~ ~ ~ ~
\ee
where $\mu_{\chi N} = m_\chi m_N/(m_\chi + m_N)$ is the DM-nucleus reduced mass and $\Delta E_e \equiv E_{e,f} - E_{e,i}$ is the energy transferred to the scattered electron.

We emphasize that these kinematics are \emph{identical} for electron scattering and Migdal scattering, provided the ionized electron is treated as a scattering state of the electron cloud Hamiltonian. In thinking of the nucleus and electrons as a single many-particle system in this formalism, it helps to regard $\vecq$ as simply the momentum transferred \emph{from} the DM, rather than as a momentum transferred \emph{to} any particular constituent of the target system. However, since the nucleus makes up the vast majority of the mass of the atom, one may think of $\vecq$ as the nuclear recoil momentum, as shown in Eq.~(\ref{eq:qcons}).

\subsection{Dynamics}
\label{sec:Dynamics}

While the kinematics of Migdal and electron scattering are identical, their dynamics differ in a crucial way depending on whether DM interacts directly with electrons or nuclei. In the language of nonrelativistic quantum mechanics, the perturbing Hamiltonian for the DM-atom interaction in the case of electron scattering is
\be
H_{{\rm int}, e} = - \int \frac{d^3 \, \mathbf{q}}{(2\pi)^3}e^{i \mathbf{q} \cdot(- \mathbf{x}_\chi + \sum_{i = 1}^{Z} \mathbf{x}_i )} \frac{\mathcal{M}_{e\chi}(q)}{4m_\chi m_e},
\ee
where $\mathcal{M}_{e\chi}(q)$ is the Lorentz-invariant matrix element for DM scattering off a free electron through 4-momentum transfer $q \approx (0, \vecq)$, and $\mathbf{x}_\chi$ and $\mathbf{x}_i$ are the position operators for the DM and electrons, respectively. Because the DM interacts directly with electrons, we can ignore the nuclear part of the atomic Hamiltonian, and the rate will be proportional to \cite{Essig:2011nj}
\be
\label{eq:Rematrix}
R_e \propto |\langle \Psi_f | H_{\rm int} | \Psi_i \rangle|^2 \sim  |\langle \psi_f | e^{i \mathbf{q} \cdot \mathbf{x}} | \psi_i \rangle|^2,
\ee
where we have made the approximation that the initial- and final-state electron cloud wavefunctions ($\Psi_i$ and $\Psi_f$, respectively) factorize such that only a single electron (with position operator $\mathbf{x}$) participates in a transition between the single-electron states $\psi_i$ and $\psi_f$. The scale of atomic wavefunctions is parametrically the size of the atom: to make this precise, we define an effective atomic radius $r_a$ as $r_a \equiv 1/|\vecq|_{\rm max}$, where $|\vecq|_{\rm max}$ is the momentum transfer at which the matrix element in  Eq.~(\ref{eq:Rematrix}) is maximized, for a given choice of initial and final states. Thus we may expect the electronic matrix element in Eq.~(\ref{eq:Rematrix}) to be as large as $\mathcal{O}(1)$ when $|\vecq| r_a = \mathcal{O}(1)$. Typical $r_a$ are on the scale of the Bohr radius $a_0$, which is $\sim 1/\keV$ in natural units, and hence $\mathbf{q} \cdot \mathbf{x} \gtrsim \mathcal{O}(1)$ where the atomic wavefunctions have support. Note that for DM heavier than 1 MeV, a momentum transfer of $1/r_a$ is always kinematically allowed, since the DM carries momentum of at least 1 keV.

For Migdal scattering, where the fundamental DM-atom interaction is with the nucleus, the interaction Hamiltonian is
\be
H_{{\rm int}, N} = - \int \frac{d^3 \, \mathbf{q}}{(2\pi)^3}e^{i \mathbf{q} \cdot(\mathbf{x}_N - \mathbf{x}_\chi)} \frac{\mathcal{M}_{N\chi}(q)}{4m_\chi m_N},
\ee
where $\mathbf{x}_N$ is the position operator for the nucleus. $H_{{\rm int}, N}$ does \emph{not} involve the electron position operators $\mathbf{x}_i$, so by itself, it cannot induce electronic transitions. However, the light-crossing time of the electron cloud is $\sim {\rm nm}/c \sim 5 \ \keV^{-1}$, so the timescale of momentum transfer to the nucleus is ``fast''  as long as the mediator mass $m_{A'}$ satisfies $m_{A'} \gg \keV$.\footnote{\label{foot:longrange}By construction, this always holds for contact interactions, $F_{\rm DM} = 1$, but it is not clear to us if the formalism developed in Ref.~\cite{Ibe:2017yqa} remains valid for ultralight mediators with $F_{\rm DM} \propto q^{-2}$, where for sufficiently small momentum transfers, the timescale for momentum transfer can be slow enough that the atomic state changes adiabatically.} In this regime, the entire atom suddenly acquires velocity $\mathbf{v}_A$ but leaves behind its stationary electrostatic potential; the electrons of the moving atom are no longer in energy eigenstates of the old electron cloud Hamiltonian. As a result, electronic transitions can arise, but not through a matrix element with the perturbing Hamiltonian $H_{{\rm int}, N}$. 

Rather, following Ref.~\cite{Ibe:2017yqa}, we construct approximate energy eigenstates of the moving atom by applying a Galilean transformation with velocity parameter $\mathbf{v}_A$, which results in a final-state atomic wavefunction containing a phase $\exp(i\sum_{i = 1}^Z \mathbf{q}_e \cdot \mathbf{x}_i)$ multiplying the full wavefunction of the atom at rest, where $\mathbf{q}_e \equiv m_e \mathbf{v}_A$.\footnote{It is important to note that $\vecq_e$ is \emph{not} to be interpreted as the momentum of the outgoing electron, but rather the effective momentum which appears in the matrix element due to the Galilean transformation to account for the atomic recoil velocity $\mathbf{v}_A$. The advantage of defining this quantity is to explicitly enable the dipole approximation of the matrix element for the masses considered here. Care should be taken not to place too much physical interpretation in this quantity, and so we translate our final result in Eq.~(\ref{eq:spectrumscale}) below back in terms of the physical momentum of the atomic system {\bf q}.} 
Consequently, as shown in Ref.~\cite{Ibe:2017yqa}, the matrix element of the nuclear wavefunction with $H_{{\rm int}, N}$ results in a factor of $\mathcal{M}_{eN}(q)$ times the overlap of the electronic wavefunctions, and thus 
\be
\label{eq:RMmatrix}
R_M \propto |\langle \Psi_{{\mathbf{v}_A}} | \Psi_i \rangle |^2  \sim  |\langle \psi_f | e^{i \mathbf{q}_e \cdot \mathbf{x}} | \psi_i \rangle|^2,
\ee
where $\Psi_{{\mathbf{v}_A}}$ is the Galilean transformation of the initial state $\Psi_i$ of the electron cloud, with velocity parameter $\mathbf{v}_A$. We have made the same approximation as in Eq.~(\ref{eq:Rematrix}) that only a single electronic transition contributes; note that $\mathbf{q}_e$ instead of $\mathbf{q}$ appears in the exponent.

By momentum conservation from Eq.~(\ref{eq:qcons}), 
\be
\mathbf{q}_e = \frac{m_e}{m_N} \mathbf{q}.
\ee
Unlike the case for electron scattering, where $\vecq \cdot \mathbf{x} \gtrsim 1$ for all DM masses greater than an MeV, $\mathbf{q}_e \cdot \mathbf{x} \ll 1$ for all sub-GeV DM because $m_e/m_N \ll 1$. Indeed, $r_a \sim 4 a_0$ for xenon, and $\mathbf{q}_e < 1/r_a$ as long as $m_\chi < 100 \ \GeV$. As $\mathbf{q}_e \to 0$, the matrix element  in Eq.~(\ref{eq:RMmatrix}) must vanish because $\psi_f$ and $\psi_i$ are energy eigenstates of the same Hamiltonian with different energy eigenvalues, by assumption. Hence the leading order term in the Taylor expansion of the exponential is linear in $\vecq_e$, and $R_M$ scales as $\mathbf{q}_e^2 = \vecq^2 (m_e/m_N)^2$. There are also additional selection rules now that the matrix element has a dipole form, $\langle \psi_f | \mathbf{x} | \psi_i \rangle$, but in general, for a given $\mathbf{q}$ and a choice of initial- and final-state wavefunctions, the ratio of Migdal and electron scattering spectra for each $i \to f$  transition scales as
\be
\label{eq:spectrumscale}
\frac{     dR_M / d\mathbf{q} }{dR_e/d\mathbf{q} } > \vecq^2 r_a^2 \left(\frac{m_e}{m_N}\right)^2 \frac{|\mathcal{M}_{\chi N}(q)/m_N|^2}{|\mathcal{M}_{\chi e}(q)/m_e|^2},
\ee
where the appearance of $r_a$ arises from the expectation value of $\mathbf{x}$, which is parametrically of order $r_a$.
We have written the above relation as an inequality because, for sufficiently large momentum transfers ($|\vecq| \gg \keV$), the exponential in Eq.~(\ref{eq:Rematrix}) will oscillate rapidly and $R_e$ will become suppressed, thereby enhancing the Migdal rate relative to the electron scattering rate. On the other hand, the Migdal spectrum may always be approximated by a dipole matrix element which scales with $\mathbf{q}^2 r_a^2$, since $\mathbf{q}_e < 1/r_a$ for all targets relevant for sub-GeV DM.
\subsection{Spectra and rates}
\label{sec:spectrarates}

To compute the ionization rate for both processes, we must integrate over the momentum transfer $\mathbf{q}$ and the DM velocity $\mathbf{v}$, and sum over the final electronic states $\psi_f$, weighted by a delta function enforcing energy conservation (3-momentum conservation has already been enforced in the definitions of $\vecq$ and $\vecq_e$ above).\footnote{Note that integrating over $\mathbf{q}$ is equivalent to integrating over the nuclear recoil energy $E_R \approx \vecq^2/(2m_N)$, since in this paper we are concerned only with the electronic energy spectrum.}  We perform the integral over $\mathbf{v}$ by approximating the DM velocity distribution as spherically symmetric, $f(\mathbf{v}) = f(v)$, such that the total rate between initial state $i$ and final state $f$ is
\be
R_{i \to f} = \frac{\rho_\chi}{m_\chi} \int d^3 v \, f(v) \, \sigma v_{i \to f},
\ee
where $\rho_\chi$ is the local DM density. For the sum over final states, we choose the normalization \cite{Essig:2011nj,Essig:2017kqs}
\be
\sum_f = \frac{1}{2}\sum_{l' m'} \int \frac{k'^3 d\ln E_e}{(2\pi)^3},
\ee
which is appropriate for scattering states in a spherically-symmetric potential which have asymptotic momentum $k' = \sqrt{2 m_e E_e}$ and angular momentum quantum numbers $l'$ and $m'$. Here, $E_e$ is the recoil energy of the ionized electron asymptotically far away from the ionized atom; from now on our final state $f$ will always be a scattered electron of energy $E_e$, and the initial state $i$ will be a bound state of (negative) energy $E_{nl}$ indexed by principal quantum number $n$ and angular momentum quantum number $l$, appropriate for a spherically-symmetric atom ignoring spin-orbit coupling and relativistic effects. The only difference between Migdal and electron scattering in the above procedure is the expression for $\sigma v _{i \to f}$.

To perform the integral over $\mathbf{q}$ and compute $\sigma v _{i \to f}$ we must specify the free-particle matrix elements. In the dark photon model, we can define a spin-averaged fiducial cross section for DM $\chi$ scattering off an isolated target $T $ with charge $|e|$ as
in Eq.~(\ref{cross-section}). For $T = p, e$, these fiducial cross sections satisfy
\be
\label{eq:sigmasdef}
\frac{\overline \sigma_e}{\mu^2_{\chi e}} = \frac{\overline \sigma_p}{\mu^2_{\chi p}},
\ee
so $\overline \sigma_e$ and $\overline \sigma_p$ are proportional by a factor which only depends on the DM mass $\chi$. To emphasize the point that $\overline \sigma_e$ and $\overline \sigma_p$ are related in this model, we shall refer to $\overline \sigma_e$ as simply $\overline \sigma$.

The fiducial cross section defined in Eq.~(\ref{cross-section}) is related to the free-particle scattering matrix element as
\be
\label{eq:Fdef}
| \mathcal{M}(q)|^2 = \frac{16 \pi m_\chi^2 m_T^2 \overline \sigma_T}{\mu^2_{\chi T}} |F_{\rm DM}(q)|^2~,
\ee
where we have assumed that the appropriate electron and DM spins have been summed and/or averaged. Here, $F_{\rm DM}(q)$ is the DM form factor which parametrizes all momentum dependence in the free-particle matrix element: if $m_{A'} \ll m_\chi v$, $F_{\rm DM}(q) \propto 1/q^2$, while if $m_{A'} \gg m_\chi v $, $F_{\rm DM}(q) = 1$. Note that in the dark photon model with equal proton and electron couplings,
\be
\frac{|\mathcal{M}_{\chi N}(q)|^2}{m_N^2} = Z^2 |F_N(q)|^2 \frac{|\mathcal{M}_{\chi e}(q)|^2}{m_e^2}
\ee
in Eq.~(\ref{eq:spectrumscale}), where $F_N$ is the form factor of the nucleus; this relation between the matrix elements gives Eq.~(\ref{eq:money}).

Putting all the pieces together, the electron recoil spectrum per unit detector mass for both electron and Migdal scattering is
\be
\frac{d R_{e,M}}{d \ln E_e} = 
N_T\frac{\rho_\chi}{m_\chi}\frac{\overline \sigma}{8 \mu_{\chi e}^2 } I_{e,M}(E_e),
\ee
where $N_T$ is the number of atomic targets and 
\be
\label{eq:Ie}
I_{e,M}(E_e) = \int d|\vecq|  |\vecq|  |F_{\rm DM}(q)|^2 \eta(\vmin) |f_{e,M}(E_e, \vecq)|^2.~~~~~~
\ee
Here, we have solved the delta function for energy conservation, $\delta(E_e - E_{nl} + \frac{\vecq^2}{2\mu_{\chi N}} - \vecq \cdot \mathbf{v})$, to perform the integral over the DM velocity distribution, resulting in a factor of $\eta(\vmin) \equiv \langle v^{-1} \theta( v- \vmin)  \rangle$, the mean inverse DM speed in the lab frame, as a function of
\be
\vmin = \frac{\Delta E_e}{|\vecq|} + \frac{|\vecq|}{2\mu_{\chi N}} = \frac{|E_{nl}| + E_e}{|\vecq|} + \frac{|\vecq|}{2\mu_{\chi N}},
\label{eq:vmin}
\ee
which is the minimum DM velocity required to ionize the target electron through a momentum transfer $|\vecq|$. The lab frame velocity distribution is cut off at $v_E + v_{\rm esc.}$, where $v_E \sim 240 \ {\rm km/s}$ is the average speed of the Earth relative to the DM halo, and $v_{\rm esc.} = 544 \ {\rm km/s}$ is the galactic escape velocity (these parameters are chosen to facilitate comparisons with Ref.~\cite{Essig:2017kqs}).

\begin{figure*}[t!]
\includegraphics[width=0.45\textwidth]{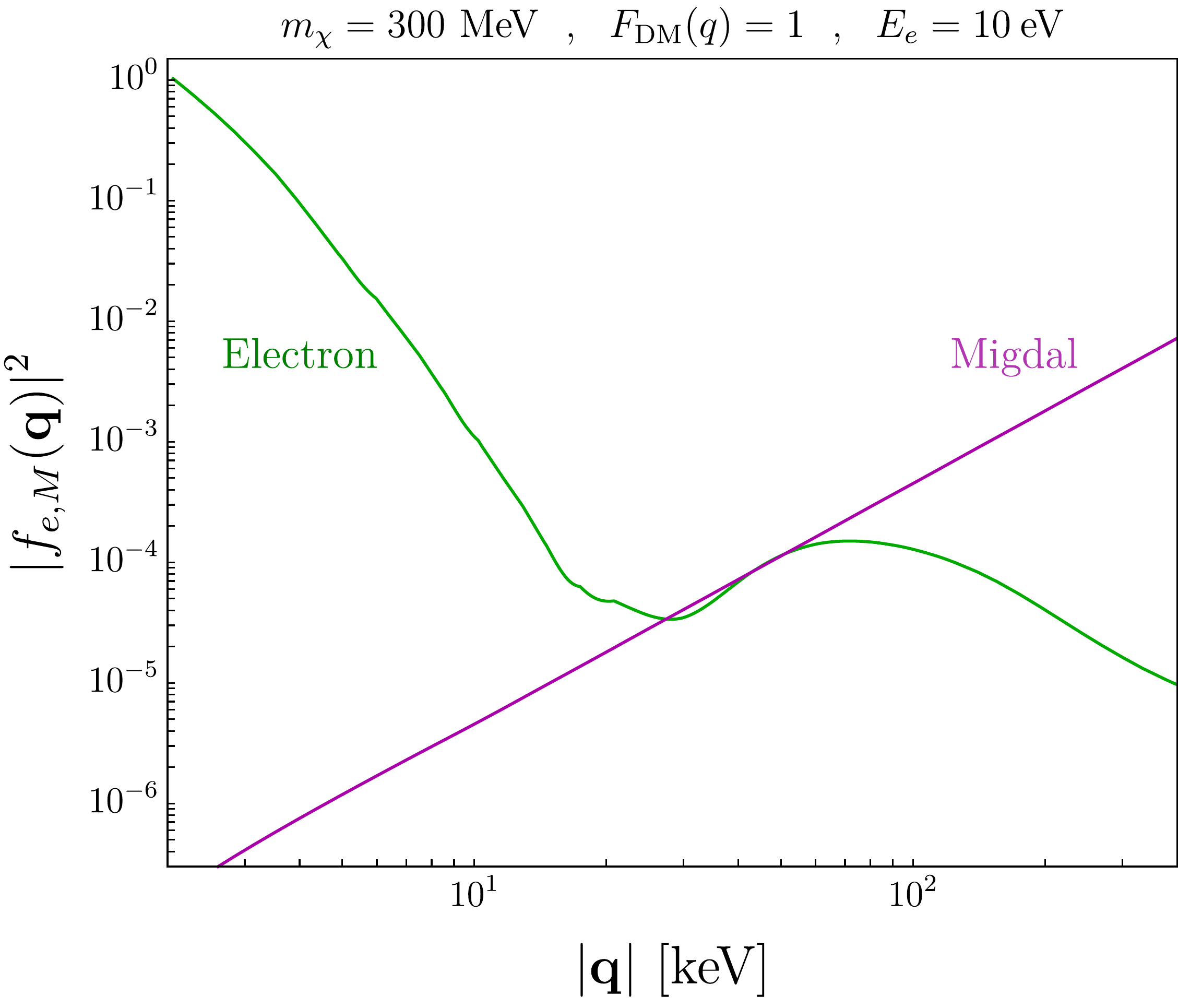}
\qquad \qquad
\includegraphics[width=0.45\textwidth]{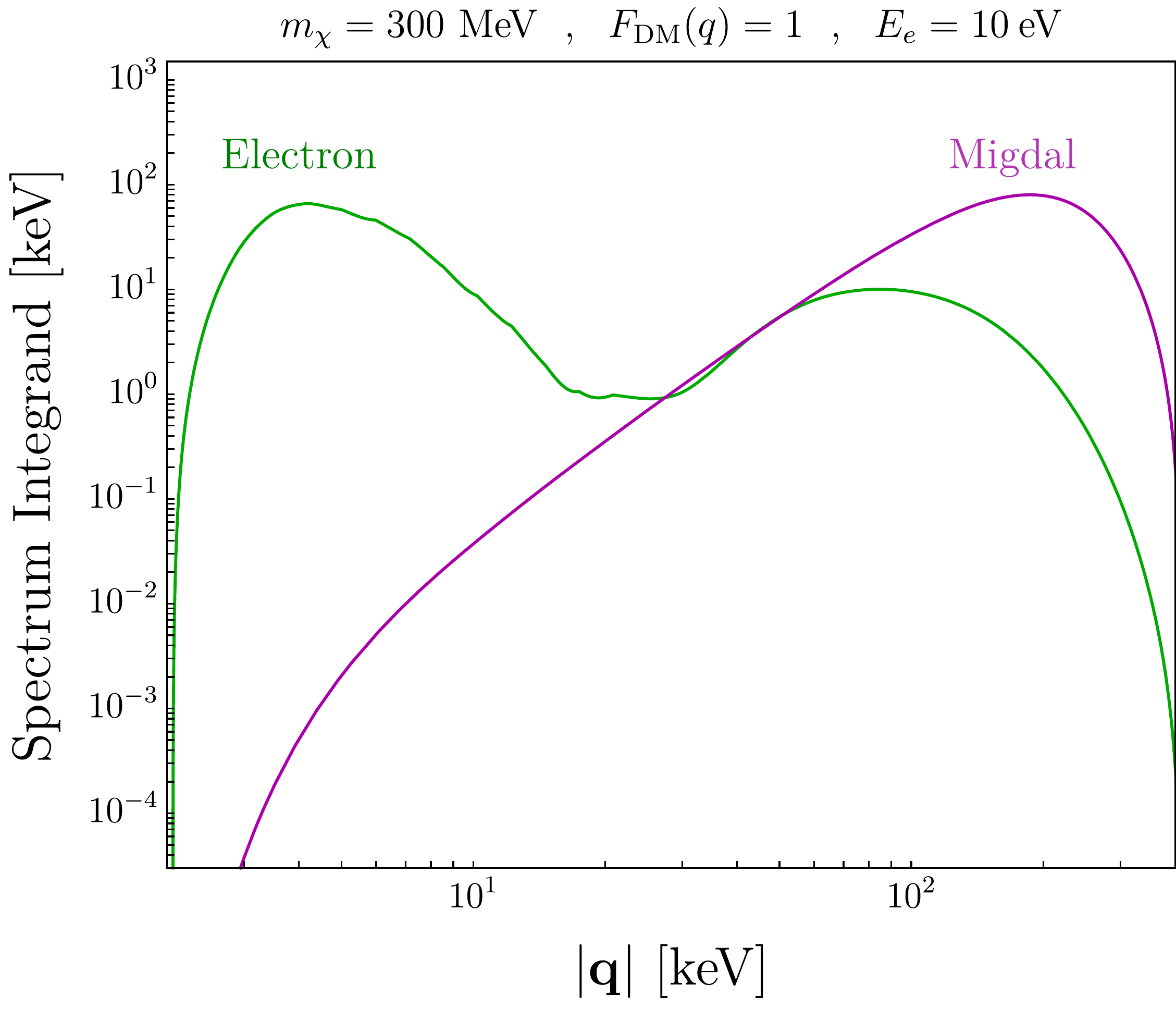}
\caption{Qualitative comparison of electron (green) and Migdal (purple) ionization form factors and spectrum integrands for $m_\chi = 300 \ \MeV$, $E_e = 10 \ \eV$, and $F_{\rm DM} = 1$. The form factors from Eqs.~(\ref{eq:fe}) and (\ref{eq:fM}) are plotted in the left panel, and the integrand from Eq.~(\ref{eq:Ie}) (which is weighted by the inverse mean DM speed) is plotted in the right panel. The electron scattering rate is dominated by small $\vecq$ while the Migdal scattering rate is dominated by large $\vecq$. These plots are computed using initial- and final-state electronic wavefunctions constructed from a hydrogenic potential for Xe, but the qualitative features are independent of the choice of wavefunctions.}
\label{fig:qcompare}
\end{figure*}

The differences between the Migdal and electron scattering processes are entirely contained in the ionization form factors $ |f_{e,M}(E_e, \vecq)|^2$, which are independent of all DM properties and depend only on the electronic and nuclear structure of the target. We will discuss in some detail in Sec.~\ref{sec:Binding} the issues with accurately computing the atomic wavefunctions required for these ionization form factors. For electron scattering,
\be
\label{eq:fe}
|f_e(E_e, \vecq)|^2 = \frac{k'^3}{4 \pi^3} \times 2 \sum_{n,l,l',m'} |\langle \psi^f_{E_e} |e^{i \mathbf{q} \cdot \mathbf{x}} | \psi^i_{E_{nl}} \rangle|^2,
\ee
where $ \psi^i_{E_{nl}}$ is a bound orbital of energy $E_{nl}$ with unit norm (the factor of 2 accounts for the approximate spin degeneracy of occupied states) and $\psi^f_{E_e}$ is an unbound electronic state of energy $E_e$ (indexed by the continuously-valued energy $E_e$ and the angular momentum quantum numbers $l'$ and $m'$), normalized to
\be
\langle \psi_{E_1,l,m} | \psi_{E_2,l',m'} \rangle = \frac{(2\pi)^3}{k_1^2} \delta(k_1 - k_2)\delta_{ll'}\delta_{mm'}~~,
\ee
where $k_{1,2} = \sqrt{2m_e E_{1,2}}$. For Migdal scattering, the analogous ionization form factor is
\begin{align}
\label{eq:fM}
|f_M(E_e, \vecq)|^2  & =  \frac{k'^3}{4\pi^3} Z^2   |F_N(q)|^2 \nonumber \\
& \times  2\sum_{n,l,l',m'} |\langle \psi^f_{E_e} |e^{i \mathbf{q}_e \cdot \mathbf{x}} | \psi^i_{E_{nl}} \rangle|^2.~~~~~~
\end{align}
The differences with respect to electron scattering are the appearance of $Z^2$ from coherent scattering off all the protons in the nucleus, a nuclear form factor $F_N$ parametrizing loss of coherence at large momentum transfers (which is largely irrelevant for the sub-MeV momentum transfers typical of sub-GeV DM), and the appearance of $\mathbf{q}_e$ instead of $\mathbf{q}$ in the matrix element between initial and final states.

The key quantity controlling the relative size of Migdal and electron scattering rates is $\vecq$. From Eq.~(\ref{eq:vmin}), the smallest allowed $|\vecq|$ is
\be 
|\vecq|_{\rm min} = \frac{E_b}{v_{\rm max}}~,
\ee
where $E_b$ is the first ionization energy (positive by convention) of the atom in question, and $v_{\rm max}$ is the largest possible DM speed, which is the Galactic escape velocity in the lab frame. Note that $|\vecq|_{\rm min}$ is \emph{independent} of the DM mass: for xenon with $E_b \sim 12 \ \eV$, and $v_{\rm max} \sim 770 \ {\rm km}/{\rm s}$, $|\vecq|_{\rm min} \sim 5 \ \keV$. Thus for all kinematically-allowed momentum transfers, $|\vecq|r_a > 1$, and electron scattering is dominated by the \emph{smallest} possible $\vecq$ before $f_e(E_e, \vecq)$ is suppressed by the quickly-oscillating exponential in the matrix element. On the other hand, the largest allowed $|\vecq|$ is
\be
|\vecq|_{\rm max} = 2 \mu_{\chi N}v_{\rm max} \sim 5 \ \keV \left( \frac{ m_\chi }{\MeV} \right),
\ee
which grows with DM mass and can be as large as hundreds of keV for $m_\chi = \mathcal{O}(100 \ \MeV)$. For these momentum transfers, $|f_M(E_e, \vecq)|^2$ still does not feel any suppression from the nuclear form factor $F_N$, which is still $\sim 1$ for $|\vecq| \lesssim \MeV$, and likewise is still in the regime of small $\vecq_e$ and so grows with $\vecq^2$. Thus the Migdal ionization form factor is largest when $\vecq$ is the largest, and Migdal scattering is dominated by the \emph{largest} kinematically-allowed momentum transfers.\footnote{In principle, there should be interference between electron and Migdal scattering, which have identical final states, but the distinct kinematics of these two processes should minimize these effects.}

We illustrate this behavior in Fig.~\ref{fig:qcompare} for $m_\chi = 300 \ \MeV$, $E_e = 10 \ \eV$, and $F_{\rm DM} = 1$. The left plot shows $|f_{e,M}(E_e, \vecq)|^2$ evaluated at $E_e = 10 \ \eV$, and the right plot shows the integrand of Eq.~(\ref{eq:Ie}) which is weighted by $\eta(\vmin)$. The $\vecq$ values plotted span the kinematically-allowed range between $|\vecq|_{\rm min}$ and $|\vecq|_{\rm max}$, as can be seen from the right plot where the velocity distribution cuts off the integrand at small and large $\vecq$. To compute $f_{e,M}$ for both electron and Migdal scattering from the same set of wavefunctions, the orthogonality of $\psi^f$ and $\psi^i$ is crucial, and to ensure this, both bound and free wavefunctions must be constructed from the same atomic Hamiltonian. A complete treatment would require a full numerical solution to the many-body Schr\"{o}dinger equation for the atom in question, but here we capture the essential features by using hydrogenic wavefunctions for the 5p shell of Xe and matching the effective nuclear charge to the binding energy of the 5p state, with scattering states constructed from the same hydrogenic potential. This unphysical choice of wavefunctions is for illustrative purposes only; the wavefunctions used in the remainder of this paper are discussed in detail in Sec.~\ref{sec:Binding}. We note that for a DM form factor proportional to $1/q^2$, as would be the case for an ultralight dark photon mediator, the spectrum integrand is weighted by $1/q^4 \sim 1/\vecq^4$ which heavily suppresses the Migdal spectrum compared to the electron spectrum for all DM masses.\footnote{As noted in Footnote~\ref{foot:longrange}, long-range interactions may result in adiabatic rather than sudden changes in atomic states during Migdal scattering, which would further suppress electronic transitions.} For these form factors, electron scattering always dominates over Migdal scattering by several orders of magnitude, and as such, for the remainder of this paper we will focus on the case $F_{\rm DM} = 1$.

\begin{figure}[t]
\hspace{2cm}
\includegraphics[width=0.47\textwidth]{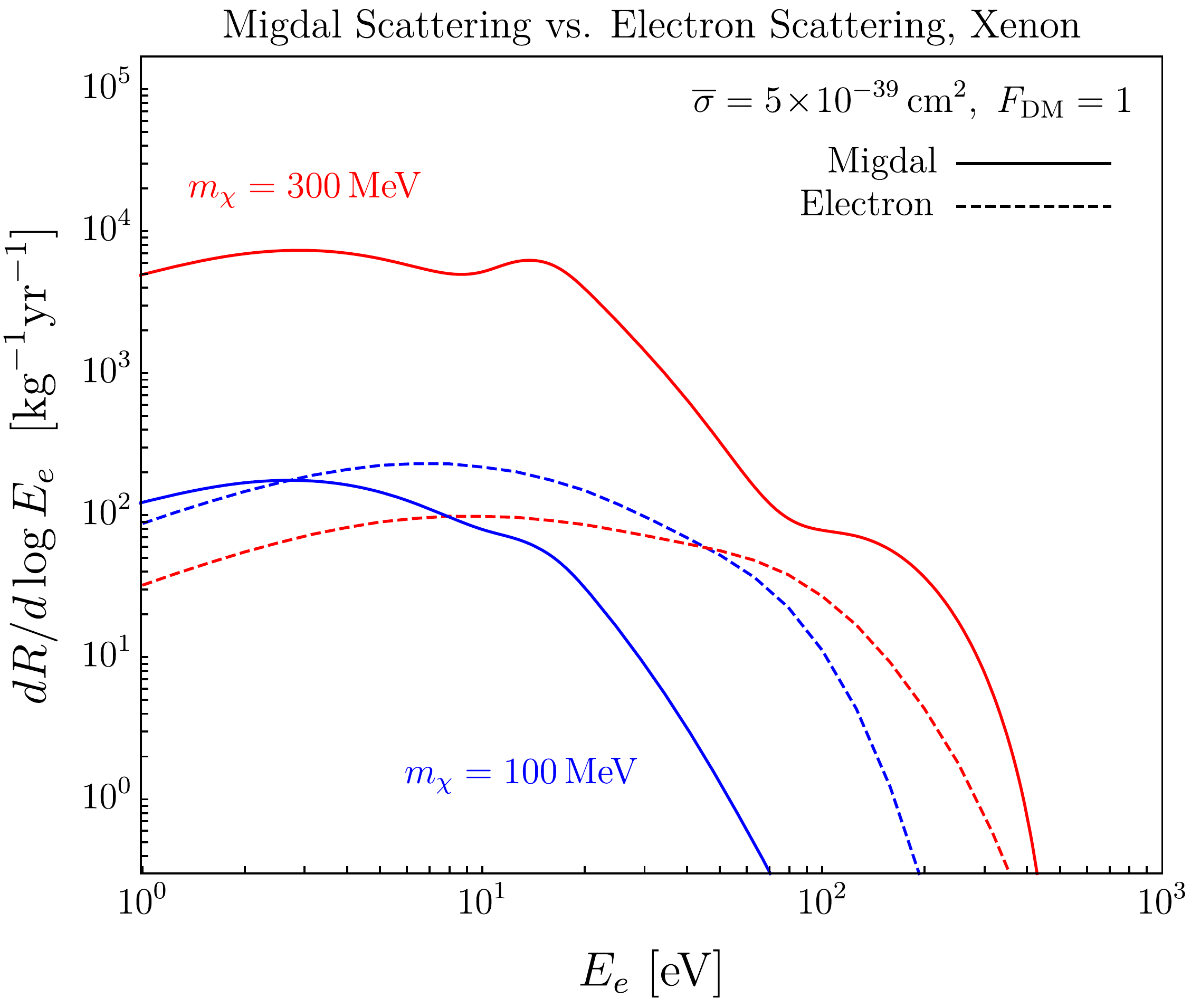}~~~~~
\caption{Comparison of electron (dashed) and Migdal (solid) spectra for reference values $\overline \sigma = 5\times10^{-39} \, \rm cm^2$ and
$m_\chi = 100$ MeV (blue) and 300 MeV (red). For both spectra, we show the inclusive rates summed
over all $E_{nl} \to E_e$ transitions in xenon, where contributions from $nl$ = 5p, 5s, and 4d dominate. Migdal spectra are computed using the wavefunctions and binding energies from Ref.~\cite{Ibe:2017yqa}, while electron spectra are computed using wavefunctions and binding energies from Ref.~\cite{Essig:2017kqs}. The differences between these choices are irrelevant for our qualitative argument here and are discussed further in Sec.~\ref{sec:Binding}.}
\label{migdal-spectrum}
\end{figure}

\begin{figure*}[!t]
\hspace{-1cm}
\includegraphics[scale=0.36]{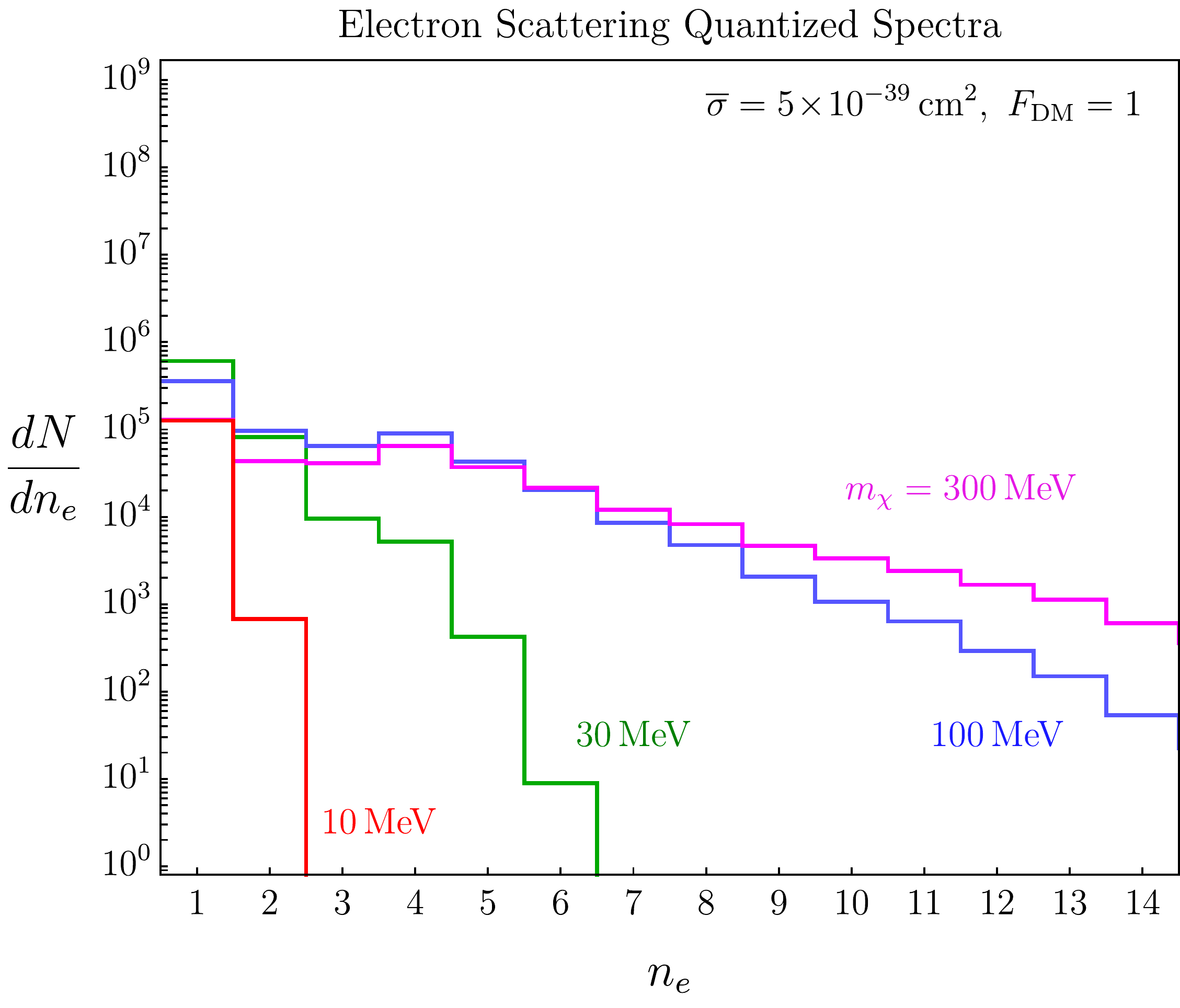}~~~~~
\includegraphics[scale=0.36]{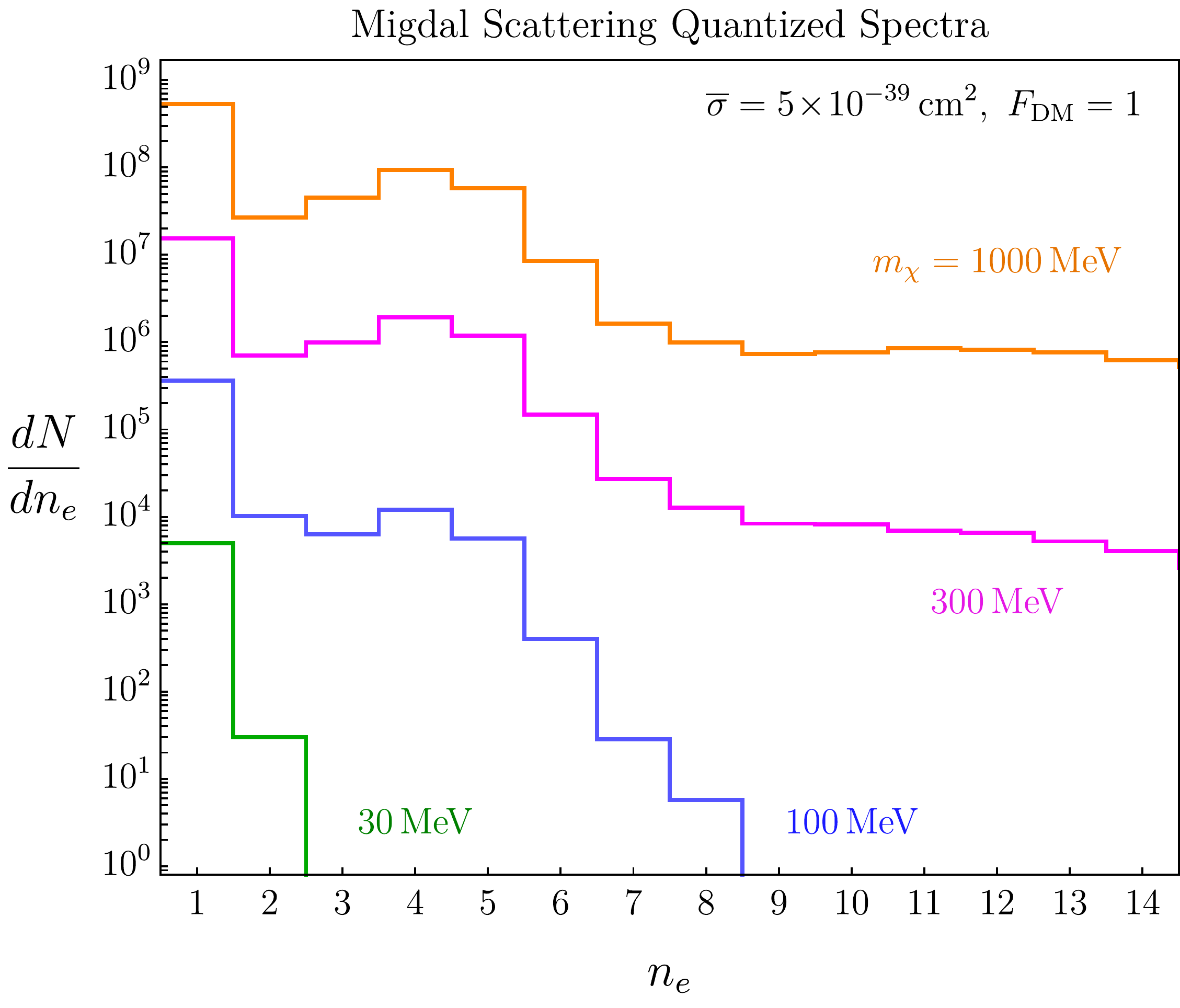} \\
\caption{Quantized rate spectra for DM-electron scattering (left) and DM-Migdal scattering (right) in xenon per number of electrons observed in the case of a heavy mediator ($F_{\rm DM} =1$) for various DM masses between 10 and 1000~MeV. We normalize these spectra to an exposure of 1000~kg-yr and a fiducial cross section of $\overline{\sigma} = 5 \times 10^{-39}$~cm$^2$.}
\label{fig:quantized_spectra}
\end{figure*}

We can confirm the relative strength of Migdal and electron scattering by considering the full electron recoil spectrum, as shown in Fig.~\ref{migdal-spectrum}. Here, to facilitate comparison with the literature, the electron spectra are calculated using the wavefunctions and binding energies of Ref.~\cite{Essig:2017kqs}, and the Migdal spectra are calculated using $f_M$ as tabulated in Ref.~\cite{Ibe:2017yqa}. Despite some differences in these wavefunctions and binding energies (which we discuss further in Sec.~\ref{sec:Binding}), the intuition developed above holds well: for sufficiently large DM masses and equal couplings to protons and electrons, the Migdal spectrum dominates over the electron spectrum for all electron recoil energies.




\section{Numerical modeling and systematic uncertainties}
\label{sec:ion}
To apply the formalism of the previous section to experimental data, we must choose a model for generating ionization spectra from recoil spectra, as well as a set of atomic wavefunctions. We choose the previously published low-threshold analysis~\cite{Essig:2017kqs} of the XENON10~\cite{angle:2011xen10,essig:2012dme1} and XENON100~\cite{aprile:2016lmdm} detectors, as well as the newly-released S2-only analysis~\cite{Aprile:2019xxb} from the XENON1T detector~\cite{Aprile:2019det}. These data are chosen for containing a relatively large exposure (for this mass range) at extremely low thresholds. All of these experiments use xenon time projection chambers to measure charge and light produced from energy deposited in liquid xenon. 

An interaction in the xenon will create some number of xenon ions $N_i$ and initially excited xenon atoms $N_{\star}$. These atoms will form dimer states in the xenon which will release energy in the form of charge and UV scintillation photons. Scintillation photons produced at this step are immediately detected in what is referred to as the S1 signal. Before they can recombine, emitted electrons are drifted in a $\sim$300~V/cm electric field to a liquid-gas interface where a stronger ($\sim$10~kV/cm) extraction field is used to accelerate the electrons into the gas, producing a second (amplified) burst of light, referred to as the S2 signal. It has been well measured that (relatively speaking) interactions with xenon nuclei preferentially deposit energy via S1, whereas interactions with electrons in the detector preferentially deposit energy via S2~\cite{dahl:2009phd,Akerib:2015wdi,Akerib:2016dd,Aprile:2018back}. In the case of sub-GeV DM interacting with a xenon atom through either electron or Migdal scattering, the momentum transfer to the recoiling atom is sufficiently small that the S1 signal is expected to be effectively zero. Thus, we consider an S2-only analysis using a 1(4) electron threshold for XENON10(100)~\cite{Essig:2017kqs} and a 5 electron threshold for XENON1T~\cite{Aprile:2019xxb}.

\subsection{Ionization model and quantization}

To compare with data, we must quantize the calculated recoil spectra in terms of the number of electrons extracted. For this, we adopt the ionization model from Refs.~\cite{Essig:2011nj,Essig:2017kqs} to determine the number of electrons produced ($n_e$) from an initial energy transfer $\Delta E_e$ which ionizes an electron from a specified electron shell with binding energy $E_{nl}$ to the continuum with energy $E_e$. We begin by considering the ejected electron, which has a probability $f_R$ to recombine (avoiding detection). According to the Thomas-Imel recombination model, $f_R$ is determined to be very small at low energies~\cite{dahl:2009phd,sorensen:2011box} in good agreement with measurement~\cite{akerib:2007xe127} and is assumed to be zero for this analysis. We can thus write the probability of observing an initially produced electron as
\be
f_0 = \frac{  1-f_R }{     1+ (N_{\star} / N_{i})  } \approx 0.83,
\ee
where the ratio of initially excited atoms to initially ionized atoms satisfies $N_{\star}/N_{i} \approx 0.2$ at high energies~\cite{doke:2002yield,aprile:2007gam}.

At high energies, the average energy $W$ required to produce one charge quantum in xenon is measured to be $W = 13.7$ eV~\cite{dahl:2009phd}. To convert $E_e$ and $E_{nl}$ into an expected quantized signal, we consider $n_t$ trials of a binomial process with probability of success, $f_0$, which satisfy
\be
n_t = {\rm floor}\left(\frac{ E_{e}}{W}\right) + {\rm floor}\left(\frac{  |E_{nl}|-E_b }{W}\right),
\ee
where $|E_{nl}|-E_b$ is the available de-excitation energy. Thus, we can write
\be
n_e = (1-f_R) +  {   {n_t}\choose{f_0}}, 
\ee
where we assume that the number $n_e$ of quanta produced is equivalent to the number of electrons extracted (observed), as the extraction efficiency should be $\sim$100$\%$~\cite{gushchin1982emission,Xu:2019ext}. Examples of quantized spectra for the $F_{\rm DM}=1$ case and different DM masses are shown in Fig.~\ref{fig:quantized_spectra}. This simple ionization model is sufficient for comparing electron and Migdal scattering here, but a more robust model would be needed to correctly interpret a signal through either channel.

Upper limits on the number of events for each value of $n_e$ in XENON10(100) have been determined in Ref.~\cite{Essig:2017kqs} to be $r_1 < 15.18$, $r_2 < 3.37$, $r_3 < 0.95$, and $r_4 < 0.35 (0.17)$ counts kg$^{-1}$ day$^{-1}$. Upper limits from XENON1T~\cite{Aprile:2019xxb} are not given directly. Instead, an upper bound of 22.5 events is reported in the range 165-275 photoelectrons in 15~ton-days of exposure\footnote{The average exposure for the range 165-198 photoelectrons has been determined from Figs. 1 and 4 of Ref.~\cite{Aprile:2019xxb}.}. We take the measured ratio of photoelectrons detected per $n_e$ (single electron gain) to be $\sim$33~\cite{Aprile:2019xxb} and conservatively assume that all events in the reported range are for the lowest encompassed bin $n_e = 5$, to obtain $r_5 < 0.0015$ counts kg$^{-1}$ day$^{-1}$. These rates already account for analysis and detector efficiencies and thus can be directly compared against our quantized spectra.

\subsection{Electron Binding Energies and Wavefunctions}
\label{sec:Binding}

 \begin{table}[t]
\begin{center}
\begin{tabular}{ c c c c c c c }
\hline 
\hline
Method & & $E_b$ & & $|E_{\rm 5p}| $ & $|E_{\rm 5s}|$ & $|E_{\rm 4d}| $ \\ \hline
Ibe et al. ~\cite{Ibe:2017yqa} & & 9.8 & & 9.8 & 21 & 61 \\
Essig et al.~\cite{Essig:2017kqs,1993ADNDT..53..113B} & & 12.4 & & 12.4 & 25.7 & 75.6 \\
Measured~\cite{brandi2001vacuum,cardona:1978xe} & & 12.1 & & 12.8 & 23.3 & 68.5 \\
\hline \hline 
\end{tabular}
\caption{Comparison of the ionization energy $E_b$ and electron binding energies $|E_{nl}|$ (eV) of the 5p, 5s, and 4d electron shells of xenon from calculations using the formalisms of Refs.~\cite{Ibe:2017yqa,Essig:2017kqs}. The measured values~\cite{brandi2001vacuum,cardona:1978xe} tend to fall somewhere in the middle.}
\label{tab:binding}
\end{center}
\end{table}

\begin{figure}[t!]
\includegraphics[scale=0.35]{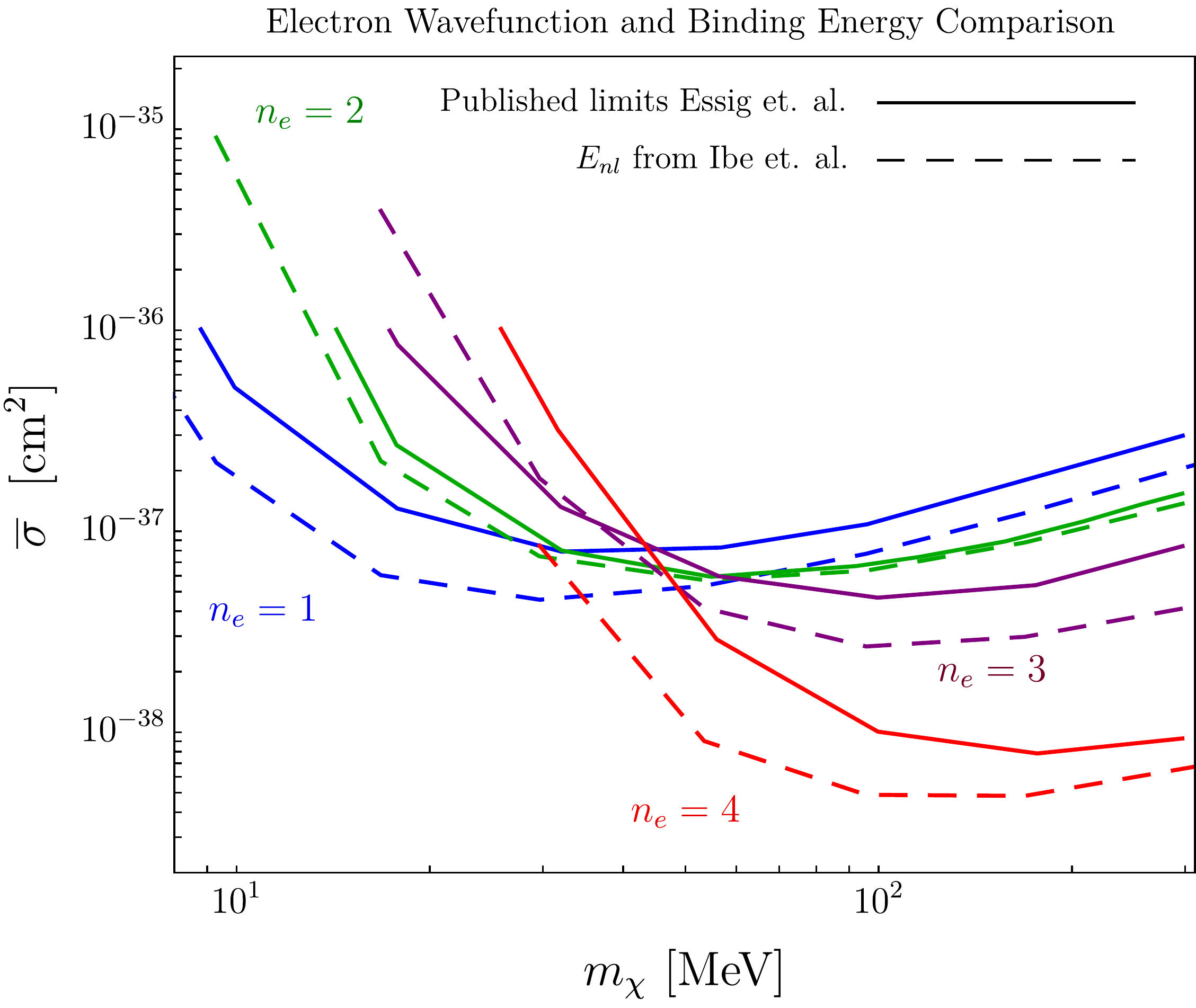}
\caption{Effect of wavefunction and binding energy choices on 90$\%$ CL limits on $\overline{\sigma}$ for direct DM-electron scattering in XENON10 with $n_e$ = 1 (blue), 2 (green), and 3 (purple) and in XENON100 with $n_e$ = 4 (red). Solid curves are published limits from Ref.~\cite{Essig:2017kqs} and dashed curves use binding energies from Ref.~\cite{Ibe:2017yqa} to construct hydrogenic final-state wavefunctions.}
\label{fig:binding_energy}
\end{figure}

The dominant quantity controlling the sensitivity for small DM masses ($\lesssim 100 \ \MeV$) is the outer-shell binding energy of xenon, or equivalently its first ionization energy. This quantity is well-measured experimentally with a value of 12.1 eV \cite{brandi2001vacuum}. Following Ref.~\cite{Ibe:2017yqa}, we define the electron shell binding energies $E_{nl}$ as an average over angular momentum states $\kappa$, 
\begin{equation}
E_{nl} = \frac{1}{2} \sum_{\kappa} \delta_{l,| \kappa+1/2|-1/2}E_{n\kappa},
\end{equation}
but define the ionization energy $E_b$ as the \emph{minimum} binding energy for all spin states in the atom. As a result, if spin-orbit coupling is ignored, $E_b = |E_{\rm 5p}|$, but the measured values show a $\sim 5 \%$ difference between the two, as can be seen in Table~\ref{tab:binding}.

Ideally, one wishes to compute the ionization spectrum using atomic wavefunctions with energy eigenvalues matching the observed binding energies. The binding energies used in the two analyses in Refs.~\cite{Essig:2017kqs,Ibe:2017yqa} are compared in Table~\ref{tab:binding}. As can be seen from this table, obtaining accurate binding energies is not entirely trivial, as the \texttt{FAC} code \cite{gu2008flexible} used in Ref.~\cite{Ibe:2017yqa} gives an outer-shell binding energy of 9.8 eV, a significant difference of 20\% from the observed value. On the other hand, the procedure used in Ref.~\cite{Essig:2017kqs} takes the calculated binding energy from the Roothaan-Hartree-Fock atomic wavefunctions tabulated in Ref.~\cite{1993ADNDT..53..113B} and constructs outgoing wavefunctions from a \emph{hydrogenic} potential with a shell-dependent effective nuclear charge which reproduces the appropriate binding energy for each electron shell. As the binding energies from Ref.~\cite{1993ADNDT..53..113B} are closer to the observed values, this procedure retains the physical binding energies at the cost of losing orthogonality between initial and final electronic states, as well as any electron correlation effects. This orthogonality is crucial in order to obtain the behavior of $f_M$ as a function of $\vecq$, as discussed above, so it is not possible to compute Migdal scattering rates using these wavefunctions. However, as noted in Sec.~\ref{sec:spectrarates}, electron scattering is dominated by the region where $\mathbf{q} \cdot \mathbf{x} \gtrsim 1$, so the form factor never probes the region where the matrix element must vanish as $\vecq \to 0$; thus, the overall kinematic features of electron scattering are sufficiently captured by this formalism.

\begin{figure*}[t!]
\hspace{-1cm}
\includegraphics[width=0.48\textwidth]{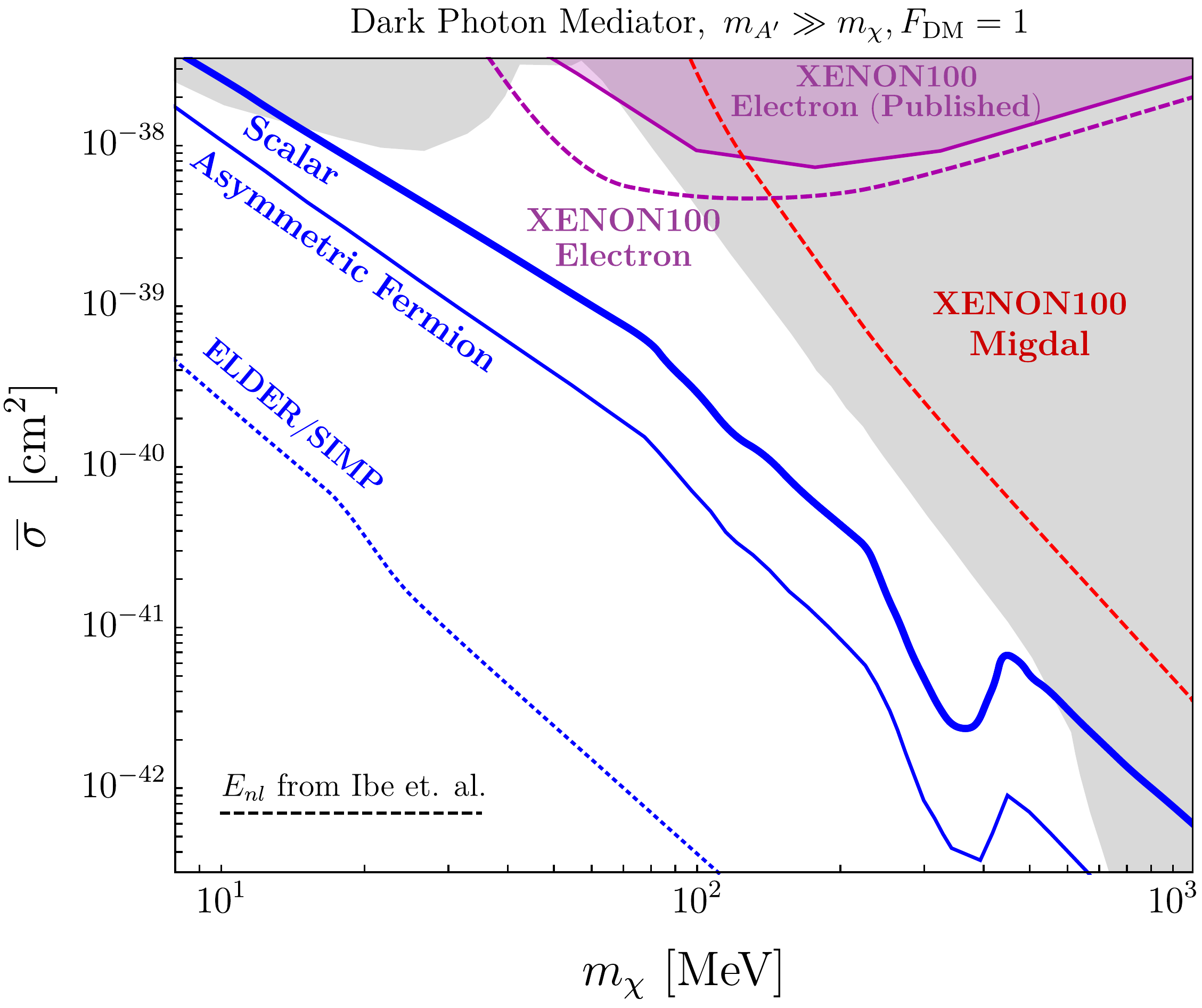}
\qquad
\includegraphics[width=0.48\textwidth]{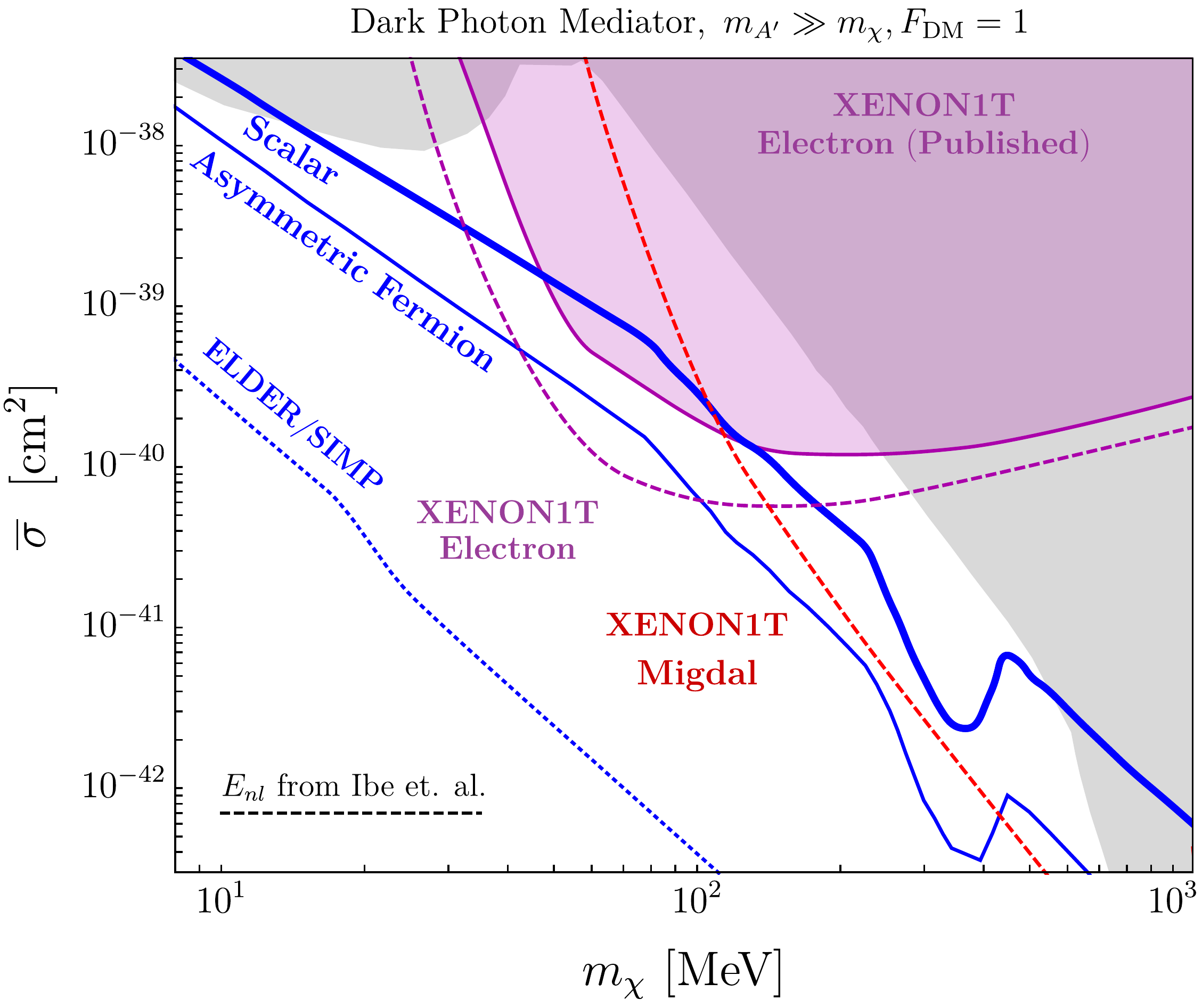}
\caption{Limits on $\overline{\sigma}$ (defined below Eq.~(\ref{eq:sigmasdef}); to compare with standard conventions, the reference cross section for electron scattering is $\overline{\sigma}_e = \sigma$, and the reference cross section for proton scattering is $\overline{\sigma}_p = \overline{\sigma} \times \mu^2_{\chi p}/\mu^2_{\chi e}$) for heavy $\apr$ ($F_{\rm DM} = 1$) mediated DM-electron (dashed purple) and DM-Migdal (dashed red) scattering are shown for $n_e=4$ for XENON100 (left) and for $n_e=5$ for XENON1T (right). In the mass and coupling range shown on the plot, the XENON10 limits are sub-dominant. For comparison, we show the published electron scattering limits~\cite{Essig:2017kqs,Aprile:2019xxb} computed with hydrogenic final-state wavefunctions and binding energies from Ref.~\cite{Essig:2017kqs} (solid purple); our electron scattering results use the smaller (unphysical) binding energies from \cite{Ibe:2017yqa} to facilitate a comparison with Migdal scattering using the same binding energies (see Sec.~\ref{sec:Binding}).
The thick blue curve is the complex scalar DM freeze-out target (particle-antiparticle symmetric DM population). Points along this curve account for the  full DM abundance as 
long as $m_{\apr} \gg m_\chi$; near resonance at $m_{\apr} \approx 2m_\chi$ this target moves down in the parameter space, but is otherwise robust \cite{Izaguirre:2015yja,Feng:2017drg}. 
The thin blue curve is the looser asymmetric Dirac fermion DM target. Any points above this line can account for the full DM abundance, but with different particle-antiparticle asymmetries~\cite{Izaguirre:2015yja,Battaglieri:2017aum}; points below this curve are excluded by Planck limits on cosmic microwave background energy injection from the annihilation of the symmetric component \cite{Aghanim:2018eyx}. 
The dotted blue curve  taken from Ref.~\cite{Battaglieri:2017aum} represents sensitivity targets for ELDER DM \cite{Kuflik:2015isi}; points above this curve correspond to 
SIMP DM models with the same $\apr$ mediator considered here  \cite{Hochberg:2014dra}.
Shaded regions represent an envelope of exclusions from beam dump searches (LSND \cite{deniverville:2011it}, E137 \cite{Bjorken:1988as,batell:2014mga}, and MiniBooNE \cite{dharmapalan:2012xp,Aguilar-Arevalo:2018wea}), nuclear recoil direct detection limits from CRESST II  \cite{Angloher:2015ewa}, and the BaBar monophoton search for invisibly decaying dark photons \cite{Lees:2017lec,izaguirre:2013uxa,essig:2013vha}.}
\label{fig:f1_limits}
\end{figure*}

While our interest in this paper is primarily a qualitative comparison of electron and Migdal scattering, we estimate one source of systematic error which can affect both processes by computing electron scattering rates using hydrogenic final-state wavefunctions (as in Ref.~\cite{Essig:2017kqs}), but constructed from the systematically-lower binding energies used in Ref.~ \cite{Ibe:2017yqa}. The main difference from the illustrative example computed in Sec.~\ref{sec:spectrarates} is that the initial-state wavefunctions are now taken from  Ref.~\cite{1993ADNDT..53..113B} instead of using a crude hydrogenic potential, in order to isolate the effects of binding energies and final-state wavefunctions. Fig.~\ref{fig:binding_energy} compares the cross section limits obtained from these binding energies to the published electron scattering limits from Ref.~\cite{Essig:2017kqs}. The systematic error on the electron scattering case associated with this procedure is less than an order of magnitude over the full DM mass range, but as expected, smaller binding energies lead to stronger cross section limits. The same procedure cannot be directly applied to Migdal scattering as the wavefunctions from~Ref.~\cite{Essig:2017kqs} are not orthogonal, but the magnitude of the difference should be comparable (about a factor of 2 for masses above 100~MeV where Migdal scattering dominates). 

We leave to future work a precise determination of experimental limits on Migdal scattering using more sophisticated quantum chemistry codes which correctly reproduce the observed binding energies
with orthogonal wavefunctions. Indeed, recent progress for electron scattering has already been made by incorporating relativistic effects and electron-electron interactions into a many-body calculation, using the observed ionization energies from photoabsorption data as a figure of merit for the quality of the wavefunctions \cite{Pandey:2018esq}; the result is that at large $n_e$, the spectrum differs significantly from that obtained with hydrogenic final-state wavefunctions, potentially affecting the limits at large DM masses by an order of magnitude. In particular, the size of relativistic effects will grow as the atomic number of the atom increases, so this may be a significant source of systematic uncertainty for Migdal scattering in xenon.




\section{Results and conclusion}
\label{sec:conclusion}
In this paper we have placed sub-GeV DM detection via electron and Migdal scattering on equivalent theoretical footing. Intriguingly, we have found that if DM couples comparably to electrons and protons through a contact interaction ($F_{\rm DM} =1)$, the Migdal rate can dominate 
for masses above $\sim$100 MeV. Thus, all existing limits for electron scattering in such models (such as dark photon-mediated scenarios), including those from XENON10, XENON100, and XENON1T~\cite{Essig:2017kqs,Aprile:2019xxb}, have omitted the {\it dominant} signal component at higher DM masses. In Fig.~\ref{fig:f1_limits}, we recalculate the full signal for DM-xenon scattering in XENON100 and XENON1T and extract upper bounds on $\overline{\sigma}$  which include both electron and Migdal scattering. It is clear that by exploiting the combination of electron and Migdal scattering, experiments with exposures and background rates comparable to XENON1T can start to probe the target parameter space for complex scalar DM freezing out through a heavy dark photon, $m_{A'} \gg m_\chi$, but as we have emphasized, a definitive conclusion requires a more careful treatment of the atomic wavefunctions than has been used previously in the literature. In the event of a signal in the DM mass range where electron and Migdal rates are within a few orders of magnitude of each other, the unique spectral shapes can be used as a discriminant, though interference effects should be carefully considered.

Although our treatment here has focused on scattering from isolated atoms, we note that additional ionization from Migdal scattering 
should also contribute in semiconductor targets (mainly Si and Ge), for which low electronic band gaps represent the next frontier
in electron-ionization direct detection. 
This additional signal channel can be probed by numerous future and ongoing experiments, including DAMIC at SNOLAB~\cite{Aguilar-Arevalo:2019damic}, SENSEI~\cite{Abramoff:2019dfb}, SuperCDMS \cite{Agnese:2018col}, and DAMIC-M \cite{Settimo:2018qcm}.
 However, a proper comparison of electron scattering and Migdal scattering in such materials
is beyond the scope of the present work and deserves a dedicated study. At a minimum, the formalism for Migdal scattering must incorporate the nontrivial harmonic potential between neighboring ions, which may result in some portion of the DM energy loss appearing as phonons rather than electronic excitations.

Despite its robust theoretical underpinnings, Migdal scattering has not yet been experimentally observed.
We emphasize that the same theoretical uncertainties which apply to Migdal scattering are present for the case of DM-electron scattering.
We take advantage of this fact to show that the systematic uncertainty in the rate calculations due to different computations of the xenon binding energies can be estimated to be a factor of a few at high masses, where the Migdal scattering rate is dominant, but other sources of theoretical uncertainty due to relativistic effects and electron correlations may be equally important.
The uncertainty due to the ionization model does not significantly affect the relative comparison of our calculated electron and Migdal scattering limits, but does explain the bulk of the difference between our limits and those in Ref.~\cite{Aprile:2019mig} when interpreted in the dark photon model.

This analysis further highlights the importance of developing low-energy calibration techniques. We have shown that Migdal and electron scattering processes probe the \emph{same} atomic wavefunctions, but in different kinematic regimes. As noted in \cite{Pandey:2018esq}, atomic many-body effects are crucial for understanding DM-atom interactions. Calibrations of both ionization processes, Migdal scattering and electron scattering, would help to resolve the theoretical uncertainty in the wavefunctions, which we believe has been substantially underappreciated by the sub-GeV DM community.

{\it Acknowledgments.} We thank Rouven Essig, Noah Kurinsky, Masahiro Ibe, Josef Pradler, Lucas Wagner, and Tien-Tien Yu for helpful conversations. We additionally thank Evan Shockley for helpful clarification regarding the recent XENON1T results. Fermilab is operated by Fermi Research Alliance, LLC, under Contract No. DE-AC02-07CH11359 with the US Department of Energy.  This work was supported in part by the Kavli Institute for Cosmological Physics at the University of Chicago through an endowment from the Kavli Foundation and its founder Fred Kavli. We thank the KICP pheno journal club for providing a stimulating environment for discussion where this work originated. This research was supported by the Munich Institute for Astro- and Particle Physics (MIAPP) of the DFG cluster of excellence ``Origin and Structure of the Universe." GK thanks MIAPP for hospitality while part of this work was completed.

\bibliography{MigdalBib2}

\end{document}